\documentclass[pra,twocolumn,preprintnumbers,amsmath,amssymb]{revtex4}
\usepackage{amsmath}
\usepackage{amssymb}
\usepackage{graphicx}
\usepackage{graphics}
\usepackage{epsfig}
\usepackage{psfrag}
\usepackage{hyperref}

\begin{document}

\title{Particle-hole fluctuations in the BCS-BEC Crossover}
\author{S. Floerchinger${}^{a}$}
\author{M. Scherer${}^{a}$}
\author{S. Diehl${}^{b}$}
\author{C. Wetterich${}^{a}$}

\affiliation{\mbox{\it ${}^a$Institut f{\"u}r Theoretische Physik,
Philosophenweg 16, D-69120 Heidelberg, Germany}\\
\mbox{\it ${}^b$Institute for Quantum Optics and Quantum
Information of the Austrian Academy of Sciences,}\\
\mbox{\it A-6020 Innsbruck, Austria}}

\begin{abstract}
The effect of particle-hole fluctuations for the BCS-BEC crossover is investigated by use of functional renormalization. We compute the critical temperature for the whole range in the scattering length $a$. On the BCS side for small negative $a$ we recover the Gorkov approximation, while on the BEC side of small positive $a$ the particle-hole fluctuations play no important role, and we find a system of interacting bosons. In the unitarity limit of infinite scattering length our quantitative estimate yields $T_c/T_F=0.264$. We also investigate the crossover from broad to narrow Feshbach resonances -- for the later we obtain $T_c/T_F=0.204$ for $a^{-1}=0$. A key ingredient for our treatment is the computation of the momentum dependent four-fermion vertex and its bosonization in terms of an effective bound-state exchange.
\end{abstract}


\maketitle

\section{Introduction}

Ultracold gases of two-component fermions near a Feshbach resonance show a smooth crossover between Bardeen-Cooper-Schrieffer (BCS) superfluidity and Bose-Einstein condensation (BEC) of molecules \cite{ALeggett80}. The quantitatively precise understanding of BCS-BEC crossover physics  is a  challenge for theory. Experimental breakthroughs as the realization of molecule condensates and the subsequent crossover to a BCS-like state of weakly attractively interacting fermions have been achieved \cite{Experiment}. Future experimental precision measurements could provide a testing ground for non-perturbative methods. An attempt in this direction are the recently published measurements of the critical temperature \cite{Luo2007} and collective dynamics \cite{Grimm2007}.

A wide range of qualitative features of the BCS-BEC crossover is already well described by extended mean-field theories which account for the contribution of both fermionic and bosonic degrees of freedom \cite{BNozieres85, SadeMelo93}. In the limit of narrow Feshbach resonances mean-field theory becomes exact \cite{Diehl:2005an, Gurarie2007}. Around this limit perturbative methods for small Yukawa couplings \cite{Diehl:2005an} can be applied. Using $\epsilon$-expansion \cite{EpsilonExpansion} or $1/N$-expansion \cite{Sachdev06} techniques one can go beyond the case of small Yukawa couplings.

Quantitative understanding of the crossover at and near the resonance has been developed through numerical calculations using various quantum Monte-Carlo (QMC) methods \cite{MonteCarlo,MonteCarloBulgac,MonteCarloBuro,MonteCarloAkki}. Computations of the complete phase diagram have been performed from functional field-theoretical techniques, in particular from $t$-matrix approaches \cite{TMatrix}, Dyson-Schwinger equations \cite{Diehl:2005an,Diehl:2005ae}, 2-Partice Irreducible methods \cite{2PI}, and renormalization-group flow equations \cite{Birse05,Diehl:2007th,Diehl:2007ri,Gubbels:2008zz}. These unified pictures of the whole phase diagram \cite{Sachdev06, TMatrix, Diehl:2005an, Diehl:2005ae, 2PI, Diehl:2007th, Diehl:2007ri, Gubbels:2008zz}, however,  do not yet reach a similar quantitative precision as the QMC calculations.

In this paper we discuss mainly the limit of broad Fesh\-bach resonances for which all thermodynamic quantities can be expressed in terms of two dimensionless parameters, namely the temperature in units of the Fermi temperature $T/T_F$ and the concentration $c=ak_F$. Here, $a$ is the scattering length and the density of atoms is used to define $k_F$ and $T_F$ via $n=k_F^3/(3\pi^2)$ and $T_F=k_F^2$. (We use natural units with $\hbar=k_B=2M=1$.) In the broad resonance regime, macroscopic observables are to a large extent independent of the concrete microscopic physical realization, a property referred to as universality \cite{Diehl:2005an, Sachdev06, Diehl:2007th}. This universality includes the unitarity regime where the scattering length diverges, $a^{-1}=0$ \cite{Ho2004}, however it is not restricted to that region.

For small and negative scattering length $c^{-1}<0, |c|\ll 1$ (BCS side), the system can be treated with perturbative methods. However, there is a significant decrease in the critical temperature as compared to the original BCS result. This was first recognized by Gorkov and Melik-Barkhudarov \cite{Gorkov}. The reason for this correction is a screening effect of particle-hole fluctuations in the medium \cite{Heiselberg}. There has been no systematic analysis of this effect in approaches encompassing the full BCS-BEC crossover so far.

In this paper, we present an approach using an exact flow equation for the ``average action'' \cite{Wetterich1993} (or ``flowing action''). We include the effect of particle-hole fluctuations and recover the Gorkov correction on the BCS side. We calculate the critical temperature for the second-order phase transition between the normal and the superfluid phase throughout the whole crossover.

We also calculate the critical temperature at the point $a^{-1}=0$ for different resonance widths $\Delta B$. As a function of the microscopic Yukawa coupling $h_\Lambda$, we find a smooth crossover between the exact narrow resonance limit and the broad resonance result. The resonance width is connected to the Yukawa coupling via $\Delta B=h_\Lambda^2/(8\pi\mu_M a_b)$ where $\mu_M$ is the magnetic moment of the bosonic bound state and $a_b$ is the background scattering length.

The paper is organized as follows. In Sect. \ref{sec:model} we explain our microscopic model and the flow equation method we use to investigate it. Section \ref{sec:ParticleHole} reviews the effect of particle-hole fluctuations on a BCS-like superfluid. Section \ref{sec:Bosonization} generalizes this to the whole BCS-BEC crossover and explains how we incorporate particle-hole fluctuations in our renormalization-group treatment using bosonization. We present our results for the critical temperature and the phase diagram in Secs. \ref{sec:CriticalTemperature} and \ref{sec:PhaseDiagram}. Finally, Sect. \ref{sec:CrossoverNarrowResonances} describes the crossover from broad to narrow resonances and we draw conclusions in Sect. \ref{sec:Conclusions}.

\section{Model and Method}\label{sec:model}
\subsection{Microscopic model}
\label{sec:microscmodel}
We investigate the functional integral for the BCS-BEC crossover of an ultracold gas of fermionic atoms near a Feshbach resonance. We start with a microscopic action including a two-component Grassmann field $\psi=(\psi_1,\psi_2)$, describing fermions in two hyperfine states. Additionally, we introduce a complex scalar field $\phi$ as the bosonic degrees of freedom. In different regimes of the crossover, it can be seen as a field describing molecules, Cooper pairs or simply an auxiliary field. Using the resulting two-channel model we can describe both narrow and broad Feshbach resonances in a unified setting. Explicitly, the microscopic action at the ultraviolet scale $\Lambda$ reads

\begin{eqnarray}
\nonumber
S[\psi, \phi] & = & \int_0^{1/T} d\tau \int d^3x{\Big \{}\psi^\dagger(\partial_\tau-\Delta-\mu)\psi\\
\nonumber
& & +\phi^*(\partial_\tau-\frac{1}{2}\Delta-2\mu+ \nu_\Lambda)\phi\\
& & - h_\Lambda(\phi^*\psi_1\psi_2+h.c.){\Big \}}\,,
\label{eqMicroscopicAction}
\end{eqnarray}
where we choose nonrelativistic natural units with $\hbar=k_B=2M=1$, with $M$ the mass of the atoms.
The system is assumed to be in thermal equilibrium, which we describe using the Matsubara formalism. In addition to the position variable $\vec{x}$, the fields depend on the imaginary time variable $\tau$ which parameterizes a torus with circumference $1/T$. The variable $\mu$ is the chemical potential. The Yukawa coupling $h$ couples the fermionic and bosonic fields. It is directly related to the width of the Feshbach resonance. The parameter $\nu$ depends on the magnetic field and determines the detuning from the Feshbach resonance. Both $h$ and $\nu$ get renormalized by fluctuations, and the microscopic values $h_\Lambda$, and $\nu_\Lambda$ have to be determined by the properties of two body scattering in vacuum. For details, we refer to \cite{Diehl:2007th, Diehl:2008}. 

More formally, the bosonic field $\phi$ appears quadratically in the microscopic action in Eq. \eqref{eqMicroscopicAction}. The functional integral over $\phi$ can be carried out. This shows that our model is equivalent to a purely fermionic theory with an interaction term
\begin{eqnarray}
\nonumber
S_{\text{int}} & = & \int_{p_1,p_2,p_1^\prime,p_2^\prime}  \left\{-\frac{h^2}{P_\phi(p_1+p_2)}\right\}\psi_1^{\ast}(p_1^\prime){\psi_1}(p_1)\\
&& \times \psi_2^{\ast}(p_2^\prime){\psi_2}(p_2)\,\delta(p_1+p_2-p_1^\prime-p_2^\prime),
\label{lambdapsieff}
\end{eqnarray}
where $p = (p_0,\vec p)$ and the classical inverse boson propagator is given by
\begin{equation}
	P_{\phi}(q)= i q_0 + \frac{\vec{q}^2}{2}+ \nu_\Lambda-2\mu\,.
\label{eq:Bosonpropagator}
\end{equation}

On the microscopic level the interaction between the fermions is described by the tree level expression
\begin{equation}
\lambda_{\psi,\text{eff}}=-\frac{h^2}{-\omega+\frac{1}{2}\vec q^2+\nu_\Lambda-2\mu}.
\end{equation}
Here, $\omega$ is the real-time frequency of the exchanged boson $\phi$. It is connected to the Matsubara frequency $q_0$ via analytic continuation $\omega=-iq_0$. Similarly, $\vec q=\vec p_1+\vec p_2$ is the center of mass momentum of the scattering fermions $\psi_1$ and $\psi_2$ with momenta $\vec p_1$ and $\vec p_2$, respectively.

In this paper we will mainly discuss the limit of broad Feshbach resonances, which is realized in current experiments, e.g. with $\mathrm{^6Li}$ and $\mathrm{^{40}K}$. This corresponds to the limit $h\to\infty$, for which the microscopic interaction becomes pointlike, with strength $-h^2/\nu_\Lambda$. For broad Feshbach resonances, one has a far going universality. Macroscopic quantities are independent of the microscopic details and can be expressed in terms of only a few parameters. In our case this is the two-body scattering length $a$ or, at finite density, the concentration $c=ak_F$, where the Fermi momentum is related to the density by $k_F=(3\pi^2 n)^{1/3}$. At nonzero temperature, an additional parameter is given by $T/T_F$, where $T_F$  is the Fermi temperature.

\subsection{Flow equations for the effective action}
The functional renormalization group connects the microphysics, as introduced in Sect \ref{sec:microscmodel}, to macrophysics and therefore to observable thermodynamics, by means of a non-perturbative flow equation. For this description, we start with a functional integral representation of the grand canonical partition function
\begin{equation}\label{eq:partfunct}
 Z = \int \mathcal{D} \tilde\chi e^{-S[\tilde\chi]}\,.
\end{equation}
Here, $\tilde\chi$ collects bosonic and fermionic degrees of freedom. In our case it can be written as a vector of the form

\begin{equation}
 \tilde\chi(x)=\left( \tilde\phi_1(x),\tilde\phi_2(x),\tilde\psi_1(x),\tilde\psi_2(x),\tilde\psi_1^{\ast}(x),\tilde\psi_2^{\ast}(x)\right)\,,
\end{equation}
where we have decomposed the complex bosonic field into its two real components,

\begin{equation}
 \tilde\phi(x) = \frac{1}{\sqrt{2}}(\tilde\phi_1(x)+i\tilde\phi_2(x))\,.
\end{equation}
We generalize equation (\ref{eq:partfunct}) by introducing a source term $\int J \tilde\chi$ and an infrared cutoff term $\Delta S_k[\tilde\chi]$, so that we obtain

\begin{equation}
 Z_k[J]=e^{W_k[J]}=\int \mathcal{D} \tilde\chi e^{-S[\tilde\chi]-\Delta S_k[\tilde\chi]+\int J \tilde\chi}\,.
\end{equation}
Here, the first equation includes a definition of the infrared regulated functional $W_k[J]$. The cutoff term $\Delta S_k[\tilde\chi]$ is quadratic in the fields. In Fourier space it reads

\begin{equation}
 \Delta S_k[\tilde\chi]=\frac{1}{2}\int \tilde\chi^{\dagger}(q)R_k(q)\tilde\chi(q)\,,
\end{equation}
where the infrared cutoff function $R_k(q)$ has the properties

\begin{eqnarray}
 R_k(q) &\rightarrow& \infty \mbox{ for } k \rightarrow \Lambda\nonumber\,,\\
 R_k(q) &\approx& k^2 \mbox{ for } \frac{|q|}{k} \rightarrow 0\nonumber\,,\\
 R_k(q) &\rightarrow& 0 \mbox{ for } \frac{k}{|q|} \rightarrow 0\,.
\end{eqnarray}
For $k=0$ the functional $Z_k[J]$ generates the $n$-point functions, while $W_k[J]$ is the generating functional for the connected correlation functions. The average action is defined as a modified Legendre transform of $W_k[J]$,

\begin{equation}
 \Gamma_k[\chi]=\left( -W_k[J]+\int J \chi\right) -\Delta S_k[\chi]\,,
\label{eq:averageaction}
\end{equation}
with $\chi=\langle\tilde\chi\rangle_J=\delta W_k/\delta J$. The average action has the important property that it interpolates between the microscopic action for $k=\Lambda$ and the quantum effective action, when $k=0$ and the cutoff is absent. For $k \neq 0$ the average action $\Gamma_k[\chi]$ is the coarse grained free energy

\begin{eqnarray}
\nonumber
 \Gamma_k[\chi] &\rightarrow& S[\chi]\quad (k\rightarrow\Lambda)\,,\\
 \Gamma_k[\chi] &\rightarrow& \Gamma[\chi]\quad(k\rightarrow 0).
\label{eq:limitseffaction}
\end{eqnarray}

The scale dependence of the average action is given by the flow equation \cite{Wetterich1993}

\begin{equation}\label{eq:flowequwett}
 \partial_k \Gamma_k[\chi]=\frac{1}{2}\mathrm{STr}\left[ \left(\Gamma_k^{(2)}+R_k \right)^{-1} \partial_k R_k\right] \,.
\end{equation}
Here, the STr operation involves an integration over momenta as well as a summation over internal indices. We employ the second functional derivative of $\Gamma_k$

\begin{equation}
 \left( \Gamma_k^{(2)}[\chi]\right)_{ij}(p_1,p_2)=\frac{\overrightarrow\delta}{\delta\chi_{i}(-p_1)}\Gamma_k[\chi]\frac{\overleftarrow\delta}{\delta\chi_{j}(p_2)}\,.
\end{equation}
Equation (\ref{eq:flowequwett}) is an exact equation and the starting point of our investigations. It is a differential equation for a functional which translates to a system of infinitely many coupled differential equations for running couplings. In perturbation theory, the solution of Eq. (\ref{eq:flowequwett}) corresponds to the computation of infinitely many Feynman diagrams.
Finding exact solutions to Eq. (\ref{eq:flowequwett}) for non-trivial theories is not possible in practice, but one can use truncations in the space of possible functionals to find approximate solutions. Such approximations do not have to rely on the existence of a small expansion parameter such as the interaction strength and they are therefore of a non-perturbative nature. For reviews of the functional renormalization group method see \cite{ReviewRG,FermionicRG,Pawlowski}.

From the full effective action $\Gamma[\chi]=\Gamma_{k=0}[\chi]$ one can derive all macroscopic properties of the system under consideration. For example, the thermodynamic properties can be obtained from the grand canonical partition function $Z$ or the corresponding grand canonical potential $\Phi_G=-T \mathrm{ln}Z$. It is related to the effective action via

\begin{equation}
 \Gamma[\chi_{eq}]=\Phi_G/T\,.
\end{equation}
Here $\chi_{eq}$ is the solution of the field equation

\begin{equation}
 \frac{\delta}{\delta\chi}\Gamma[\chi]{\bigg |}_{\chi=\chi_{eq}}=0\,.
\end{equation}
Since the effective action is the generating functional of the one-particle-irreducible (1PI) correlation functions, we can also derive dynamical properties from $\Gamma[\chi]$. For example, for a homogeneous field $\chi_\text{eq}$, the full propagator $G(q)$ is obtained from
\begin{equation}
(\Gamma^{(2)})_{ij}(p_1,p_2)=(G^{-1})_{ij}(p_1)\, \delta(p_1-p_2).
\end{equation}

In this work, we solve the flow equation \eqref{eq:flowequwett} approximately by using a truncation in the space of possible functionals $\Gamma_k$. More explicitly, our truncation reads
\begin{eqnarray}
\nonumber
\Gamma_k[\chi] & =  & \int_0^{1/T}d\tau \int d^3x {\bigg \{} \psi^\dagger (\partial_\tau -\Delta -\mu) \psi\\
\nonumber
& + & \bar{\phi}^*(\bar Z_\phi \partial_\tau-\frac{1}{2}\bar A_\phi \Delta)\bar\phi + \bar U(\bar \rho,\mu) \\
& - & \bar h (\bar \phi^* \psi_1\psi_2 + \bar \phi \psi_2^\ast \psi_1^\ast) {\bigg \} }.
\label{eq:baretruncation}
\end{eqnarray}
Here the effective potential $\bar U(\bar \rho,\mu)$ contains no derivatives and is a function of $\bar{\rho}=\bar{\phi}^*\bar{\phi}$ and $\mu$. Besides the couplings parameterizing $\bar U$ (see below) our truncation contains three further $k$-dependent (``running'') couplings $\bar A_\phi$, $\bar Z_\phi$ and $\bar h$. The truncation in Eq. \eqref{eq:baretruncation} can be motivated by a systematic derivative expansion and analysis of symmetry constraints (Ward identities), see \cite{Diehl:2007th, Floerchinger2008b, Diehl:2008}. The truncation in Eq. \eqref{eq:baretruncation} does not yet incorporate the effects of particle-hole fluctuations and we will come back to this issue in Sect. \ref{sec:Bosonization}, Eq. \eqref{eq:fourfermionvertex}. In terms of renormalized fields $\phi=\bar A_\phi^{1/2}\bar\phi$, $\rho=\bar A_\phi \bar \rho$, renormalized couplings $Z_\phi=\bar Z_\phi / \bar A_\phi$, $h=\bar h/\sqrt{\bar A_\phi}$ and effective potential $U(\rho,\mu)=\bar U(\bar \rho, \mu)$, Eq. \eqref{eq:baretruncation} reads
\begin{eqnarray}
\nonumber
\Gamma_k[\chi] & = & \int_0^{1/T}d\tau \int d^3x {\bigg \{}  \psi^\dagger (\partial_\tau -\Delta -\mu) \psi\\
\nonumber
& + & \phi^*( Z_\phi \partial_\tau-\frac{1}{2} \Delta)\phi +  U(\rho,\mu) \\
& - &  h \,(\phi^* \psi_1\psi_2 + \phi \psi_2^\ast \psi_1^\ast) {\bigg \} }.
\label{eq:truncation}
\end{eqnarray}
For the effective potential, we use an expansion around the $k$-dependent location of the minimum $\rho_0(k)$ and the $k$-independent value of the chemical potential $\mu_0$ that corresponds to the physical particle number density $n$. We determine $\rho_0(k)$ and $\mu_0$ by the requirements 
\begin{eqnarray}
\nonumber
(\partial_\rho U)(\rho_0(k),\mu_0)=0 &&\text{for all }k\\
-(\partial_\mu U) (\rho_0,\mu_0)=n && \text{at }k=0.  
\end{eqnarray}
More explicitly, we employ a truncation for $U(\rho,\mu)$ of the form
\begin{eqnarray}
\nonumber
U(\rho,\mu) &=& U(\rho_0,\mu_0)-n_k (\mu-\mu_0)\\
&& +(m^2+\alpha(\mu-\mu_0)) (\rho-\rho_0)\nonumber\\
&& +\frac{1}{2}\lambda (\rho-\rho_0)^2.
\end{eqnarray}
In the symmetric or normal gas phase, we have $\rho_0=0$, while in the regime with spontaneous symmetry breaking, we have $m^2=0$. The atom density $n=-\partial U/\partial \mu$ corresponds to $n_k$ in the limit $k\to 0$.

In total, we have the running couplings $m^2(k)$, $\lambda(k)$, $\alpha(k)$, $n_k$, $Z_\phi(k)$ and $h(k)$. (In the phase with spontaneous symmetry breaking $m^2$ is replaced by $\rho_0$.)  In addition, we need the anomalous dimension $\eta=-k\partial_k \text{ln} \bar A_\phi$. We project the flow of the average action $\Gamma_k$  on the flow of these couplings by taking appropriate (functional) derivatives on both sides of Eq. \eqref{eq:flowequwett}.  We thereby obtain a set of coupled nonlinear differential equations which can be solved numerically. 

At the microscopic scale $k=\Lambda$ the initial values of our couplings are determined from Eq. \eqref{eq:limitseffaction}. This gives $m^2(\Lambda)=\nu_\Lambda-2\mu_0$, $\rho_0=0$, $\lambda(\Lambda)=0$, $Z_\phi(\Lambda)=1$, $h(\Lambda)=h_\Lambda$, $\alpha(\Lambda)=-2$ and $n_\Lambda=3\pi^2 \mu_0 \theta(\mu_0)$. The initial values $\nu_\Lambda$ and $h_\Lambda$ can be connected to the two particle scattering in vacuum close to a Feshbach resonance. For this purpose one follows the flow of $m^2(k)$ and $h(k)$ in vacuum, i.e. $\mu_0=T=n=0$ and extracts the renormalized parameters $m^2=m^2(k=0)$, $h=h(k=0)$. The scattering length $a$ obeys $a=-h^2/(8\pi m^2)$ and the renormalized Yukawa coupling $h$ determines the width of the resonance as discussed in Sect. \ref{sec:CrossoverNarrowResonances}. Broad Feshbach resonances with large $h$ become independent of $h$.

The infrared cutoff we use is purely space-like and reads in terms of the bare fields
\begin{eqnarray}
\nonumber
\Delta S_k &=& \int_p {\bigg \{} \psi^\dagger(p)\left(\text{sign}(\vec p^2-\mu)k^2-(\vec p^2-\mu)\right) \\
\nonumber
&& \times \quad \theta\left(k^2-|\vec p^2-\mu|\right)\psi(p)\\
\nonumber
&& + \bar \phi^*(p)\bar A_\phi \left(k^2-\vec p^2/2\right)\theta\left(k^2-\vec p^2/2\right) \bar \phi(p) {\bigg \}}.\\
\label{eq:cutoff}
\end{eqnarray}
For the fermions it regularizes fluctuations around the Fermi surface, while for the bosons fluctuations with small momenta are suppressed. The choice of $\Delta S_k$ in Eq. \eqref{eq:cutoff} is an optimized choice in the sense of \cite{Litim,Pawlowski}.

The flow equations without the effect of the particle-hole exchange have been discussed extensively elsewhere \cite{Diehl:2007th, Diehl:2007ri}. Even though we use here a different cutoff and a different treatment of the density, we will not give them explicitly and rather concentrate on the effects of particle-hole exchange diagrams.

\section{Particle-hole fluctuations}\label{sec:ParticleHole}

The BCS theory of superfluidity in a Fermi gas of atoms is valid for a small attractive interaction between the fermions \cite{BCS}. In a renormalization group setting, the features of BCS theory can be described in a purely fermionic language. The only scale dependent object is the fermion interaction vertex $\lambda_\psi$. The flow depends on the temperature and the chemical potential. 
For positive chemical potential ($\mu>0$) and small temperatures $T$, the appearance of pairing is indicated by the divergence of $\lambda_\psi$.

\begin{figure}[ht]
\centering
\includegraphics[width=0.45\textwidth]{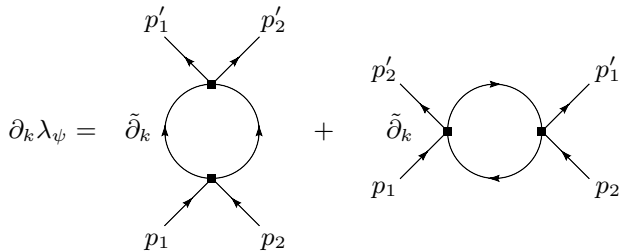}
\caption{Running of the momentum dependent vertex $\lambda_{\psi}$. Here $\tilde{\partial}_k$ indicates derivatives with respect to the cutoff terms in the propagators and does not act on the vertices in the depicted diagrams. We will refer to the first loop as the particle-particle loop (pp-loop) and to the second one as the particle-hole loop (ph-loop).}
\label{fig:lambdaflow}
\end{figure}

In general, the interaction vertex is momentum dependent and represented by a term
\begin{eqnarray}
\Gamma_{\lambda_\psi}&=&\int_{p_1,p_2,p_1^\prime,p_2^\prime}\lambda_{\psi}(p_1^\prime,p_1,p_2^\prime,p_2)\nonumber\\
& &\times\psi_1^{\ast}(p_1^\prime)\psi_1(p_1)\psi_2^{\ast}(p_2^\prime)\psi_2(p_2)
\label{eq:momentumdepvertex}
\end{eqnarray}
in the effective action. In a homogeneous situation, momentum conservation restricts the expression in Eq. \eqref{eq:momentumdepvertex} to three independent momenta, $\lambda_{\psi}\sim \delta(p_1^\prime+p_2^\prime-p_1-p_2)$. The flow of $\lambda_\psi$ has two contributions which are depicted in Fig. \ref{fig:lambdaflow}. 
The first diagram describes particle-particle fluctuations. For $\mu>0$ its effect increases as the temperature $T$ is lowered. For small temperatures $T\leq T_{c,\text{BCS}}$ the logarithmic divergence leads to the appearance of pairing, as $\lambda_\psi\to \infty$. 

In the purely fermionic formulation the flow equation for $\lambda_{\psi}$ has the general form \cite{FermionicRG, Ellwanger}

\begin{equation}\label{eq:lambdapsi2}
\partial_k \lambda_{\psi}^{\alpha}=A^{\alpha}_{\beta\gamma}\lambda_{\psi}^{\beta}\lambda_{\psi}^{\gamma}\,, 
\end{equation}
with $\alpha, \beta, \gamma$ denoting momentum as well as spin labels. A numerical solution of this equation is rather involved due to the rich momentum structure. The case of the attractive Hubbard model in two dimensions, which is close to our problem, has recently been discussed in \cite{Strack}.
The BCS approach concentrates on the pointlike coupling, evaluated by setting all momenta to zero. For $k \rightarrow 0,\ \mu_0 \rightarrow 0,\ T \rightarrow 0$ and $n \rightarrow 0$ this coupling is related the scattering length, $a= \frac{1}{8\pi} \lambda_{\psi}(p_i=0)$. In the BCS approximation only the first diagram in Fig. \ref{fig:lambdaflow} is kept, and the momentum dependence of the couplings on the right-hand side of Eq. \eqref{eq:lambdapsi2} is neglected, by replacing $\lambda_{\psi}^{\alpha}$ by the pointlike coupling evaluated at zero momentum. In terms of the scattering length $a$, Fermi momentum $k_F$ and Fermi temperature $T_F$, the critical temperature is found to be 
\begin{equation}
 \frac{T_c}{T_F}\approx 0.61 e^{\pi/(2 a k_F)}\,.
\end{equation}
This is the result of the original BCS theory. However, it is obtained by entirely neglecting the second loop in Fig. \ref{fig:lambdaflow}, which describes particle-hole fluctuations. At zero temperature the expression for this second diagram vanishes if it is evaluated for vanishing external momenta. Indeed, the two poles of the frequency integration are always either in the upper or lower half of the complex plane and the contour of the frequency integration can be closed in the half plane without poles. 

The dominant part of the scattering in a fermion gas occurs, however, for momenta on the Fermi surface  rather than for zero momentum. For non-zero momenta of the "external particles" the second diagram in Fig. \ref{fig:lambdaflow} - the particle-hole channel - makes an important contribution. 

Setting the external frequencies to zero, we find that the inverse propagators in the particle-hole loop are 
\begin{equation}\label{eq:loopmom1}
P_\psi(q)=i q_0 +(\vec{q}-\vec{p}_1)^2-\mu\,,
\end{equation}
and 
\begin{equation}\label{eq:loopmom2}
P_\psi(q)=i q_0 +(\vec{q}-\vec{p}_2^{\,\prime})^2-\mu.
\end{equation}
Depending on the value of the momenta $\vec{p}_1$ and $\vec p_2^{\,\prime}$, there are now values of the loop momentum $\vec q$ for which the poles of the frequency integration are in different half planes so that there is a nonzero contribution even for $T=0$.

To include the effect of particle-hole fluctuations one could try to take the full momentum dependence of the vertex $\lambda_\psi$ into account. However, this leads to complicated expressions which are hard to solve even numerically. 
One therefore often restricts the flow to the running of a single coupling $\lambda_\psi$ by choosing an appropriate projection prescription to determine the flow equation. In the purely fermionic description with a single running coupling $\lambda_\psi$, this flow equation has a simple structure. The solution for $\lambda_{\psi}^{-1}$ can be written as a contribution from the particle-particle (first diagram in Fig. \ref{fig:lambdaflow}, pp-loop) and the particle-hole (second diagram, ph-loop) channels 
\begin{equation}\label{PHComp}
 \frac{1}{\lambda_{\psi}(k=0)}=\frac{1}{\lambda_{\psi}(k=\Lambda)} + \mbox{pp-loop} + \mbox{ph-loop}\,.
\end{equation}
Since the ph-loop depends only weakly on the temperature, one can evaluate it at $T=0$ and add it to the initial value $\lambda_\psi(k=\Lambda)^{-1}$. Since $T_c$ depends exponentially on the "effective microscopic coupling"
\begin{equation}
 \left(\lambda_{\psi,\Lambda}^{\text{eff}}\right)^{-1}=\lambda_{\psi}(k=\Lambda)^{-1} + \text{ph-loop}\,,
\end{equation}
any shift in $\left(\lambda_{\psi,\Lambda}^{\text{eff}}\right)^{-1}$ results in a multiplicative factor for $T_c$. The numerical value of the ph-loop and therefore of the correction factor for $T_c/T_F$ depends on the precise projection description.

Let us now choose the appropriate momentum configuration. For the formation of Cooper pairs, the  relevant momenta lie on the Fermi surface, 

\begin{equation}\label{absmom}
\vec{p}^2_1=\vec{p}^2_2=\vec{p}^{{\,\prime}2}_1=\vec{p}^{{\,\prime}2}_2=\mu\,,
\end{equation}
and point in opposite directions

\begin{equation}\label{oppmom}
 \vec{p}_1=-\vec{p}_2,\ \vec{p}^{\,\prime}_1=-\vec{p}^{\,\prime}_2\,.
\end{equation}
This still leaves the angle between $\vec{p}_1$ and $\vec{p}^{\,\prime}_1$ unspecified. Gorkov's approximation uses Eqs. \eqref{absmom} and \eqref{oppmom} and projects on the $s$-wave by averaging over the angle between $\vec{p}_1$ and $\vec{p}^{\,\prime}_1$. One can shift the loop momentum such that the internal propagators depend on $\vec{q}^2$ and $(\vec{q}+\vec{p}_1-\vec{p}^{\,\prime}_1)^2$. In terms of spherical coordinates the first propagator depends only on the magnitude of the loop momentum $q^2=\vec{q}^2$, while the second depends additionally on the transfer momentum $\tilde{p}^2=\frac{1}{4}(\vec{p}_1-\vec{p}^{\,\prime}_1)^2$ and the angle $\alpha$ between $\vec{q}$ and $(\vec{p}_1-\vec{p}^{\,\prime}_1)$,

\begin{equation}
 (\vec{q}+\vec{p}_1-\vec{p}^{\,\prime}_1)^2=q^2+4\tilde{p}^2+4\,q\, \tilde{p}\,\text{cos}(\alpha)\,.
\end{equation}
Performing the loop integration involves the integration over $q^2$ and the angle $\alpha$. The averaging over the angle between $\vec{p}_1$ and $\vec{p}_1^{\,\prime}$ translates to an averaging over $\tilde{p}^2$. Both can be done analytically \cite{Heiselberg} for the fermionic particle-hole diagram and the result gives the well-known Gorkov correction to BCS theory, resulting in

\begin{equation}
T_c=\frac{1}{(4e)^{1/3}}T_{c,\text{BCS}}\approx \frac{1}{2.2} T_{c,\text{BCS}}\,.
\end{equation}

In this paper we will use a numerically simpler projection by choosing $\vec{p}^{\,\prime}_1=\vec{p}_1$, and $\vec{p}_2=\vec{p}^{\,\prime}_2$, without an averaging over the angle between $\vec{p}^{\,\prime}_1$ and $\vec{p}_1$. The size of $\tilde p^2 = \vec{p}^2_1$ is chosen such that the one-loop result reproduces exactly the result of the Gorkov correction, namely $\tilde p = 0.7326 \sqrt{\mu}$. Choosing different values of $\tilde p$ demonstrates the dependence of $T_c$ on the projection procedure and may be taken as an estimate for the error that arises from the limitation to one single coupling $\lambda_{\psi}$ instead of a momentum dependent function.

\section{Bosonization}\label{sec:Bosonization}

In Sec. \ref{sec:model} we describe an effective four-fermion interaction by the exchange of a boson. In this picture the phase transition to the superfluid phase is indicated by the vanishing of the bosonic  ``mass term'' $m^2 = 0$. Negative $m^2$ leads to the spontaneous breaking of U(1)-symmetry, since the minimum of the effective potential occurs for a nonvanishing superfluid density $\rho_0>0$.

\begin{figure}[ht]
\centering
\includegraphics[width=0.35\textwidth]{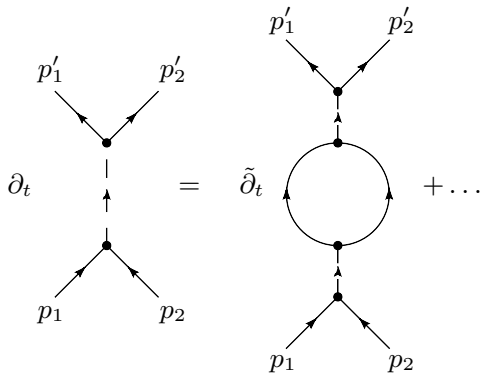}
\caption{Flow of the boson propagator.}
\label{fig:bosonexchangeloop}
\end{figure}
For $m^2 \geq 0$ we can solve the field equation for the boson $\phi$ as a functional of $\psi$ and insert the solution into the effective action. This leads to an effective four-fermion vertex describing the scattering $\psi_1(p_1)\psi_2(p_2)\to \psi_1(p_1^{\,\prime})\psi_2(p_2^{\,\prime})$
\begin{equation}
\lambda_{\psi,\text{eff}}=\frac{-h^2}{i(p_1+p_2)_0+\frac{1}{2}(\vec p_1+\vec p_2)^2+m^2}.
\label{eq:lambdapsieff}
\end{equation}
To investigate the breaking of U(1) symmetry and the onset of superfluidity, we first consider the flow of the bosonic propagator, which is mainly driven by the fermionic loop diagram. For the effective four-fermion interaction this accounts for the particle-particle loop (see r.h.s. of Fig. \ref{fig:bosonexchangeloop}). In the BCS limit of a large microscopic $m_\Lambda^2$ the running of $m^2$ for $k\to0$ reproduces the BCS result \cite{BCS}.

The particle-hole fluctuations are not accounted for by the renormalization of the boson propagator. Indeed, we have neglected so far that a term

\begin{equation}\label{eq:fourfermionvertex}
 \int_{\tau,\vec{x}}\lambda_{\psi}\psi_1^{\ast}\psi_1\psi_2^{\ast}\psi_2\,,
\end{equation}
in the effective action is generated by the flow. This holds even if the microscopic pointlike interaction is absorbed by a Hubbard-Stratonovich transformation into an effective boson exchange such that $\lambda_\psi(\Lambda)=0$. The strength of the total interaction between fermions

\begin{equation}\label{eq:lambdapsieff2}
\lambda_{\psi,\text{eff}}=\frac{-h^2}{i(p_1+p_2)_0+\frac{1}{2}(\vec p_1+\vec p_2)^2+m^2} + \lambda_{\psi}
\end{equation}
adds $\lambda_\psi$ to the piece generated by boson exchange. 
\begin{figure}[ht]
\centering
\includegraphics[width=0.35\textwidth]{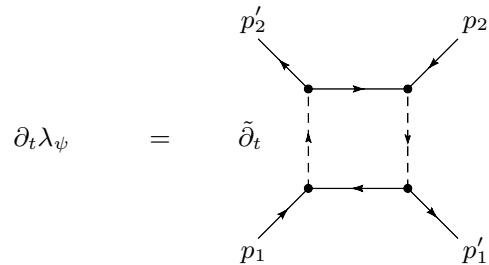}
\caption{Box diagram for the flow of the four-fermion interaction.}
\label{fig:boxes}
\end{figure}
In the partially bosonized formulation, the flow of $\lambda_\psi$ is generated by the box-diagrams depicted in Fig. \ref{fig:boxes}. We may interpret these diagrams and establish a direct connection to the particle-hole diagrams depicted in Fig. \ref{fig:lambdaflow} on the BCS side of the crossover and in the microscopic regime. There the boson gap $m^2$ is large. In this case, the effective fermion interaction in Eq. \eqref{eq:lambdapsieff2} becomes momentum independent. Diagrammatically, this is represented by contracting the bosonic propagator. One can see, that the box-diagram in Fig. \ref{fig:boxes} is then equivalent to the particle-hole loop investigated in Sec. \ref{sec:ParticleHole} with the pointlike approximation $\lambda_{\psi,\text{eff}}\to-\frac{h^2}{m^2}$ for the fermion interaction vertex. As mentioned above, these contributions vanish for $T=0$, $\mu<0$ for arbitrary $\vec p_i$. Indeed, at zero temperature, the summation over the Matsubara frequencies becomes an integral. All the poles of this integration are in the upper half of the complex plane and the integration contour can be closed in the lower half plane. We will evaluate $\partial_k \lambda_\psi$ for $\vec p_1=\vec p_1^{\,\prime}=-\vec p_2=-\vec p_2^{\,\prime}$, $|\vec p_1|=\tilde p = 0.7326 \sqrt{\mu}$, as discussed in the Sec. \ref{sec:ParticleHole}. For $\mu>0$ this yields a nonvanishing flow even for $T=0$.

Another simplification concerns the temperature dependence. While the contribution of particle-particle diagrams becomes very large for small temperatures, this is not the case for particle-hole diagrams. For nonvanishing density and small temperatures, the large effect of particle-particle fluctuations leads to the spontaneous breaking of the U(1) symmetry and the associated superfluidity. In contrast, the particle-hole fluctuations lead only to quantitative corrections and depend only weakly on temperature. This can be checked explicitly in the pointlike approximation, and holds not only in the BCS regime where $T/\mu \ll 1$, but also for moderate $T/\mu$ as realized at the critical temperature in the unitary regime. We can therefore evaluate the box-diagrams in Fig. \ref{fig:lambdaflow} for zero temperature. We note that an implicit temperature dependence, resulting from the couplings parameterizing the boson propagator, is taken into account.

After these preliminaries, we can now incorporate the effect of particle-hole fluctuations in the renormalization group flow. A first idea might be to include the additional term \eqref{eq:fourfermionvertex} in the truncation and to study the effects of $\lambda_{\psi}$ on the remaining flow equations. On the initial or microscopic scale one would have $\lambda_{\psi}=0$, but it would then be generated by the flow. This procedure, however, has several shortcomings. First, the appearance of a local condensate would now be indicated by the divergence of the effective four-fermion interaction

\begin{equation}
 \lambda_{\psi,\text{eff}}=-\frac{h^2}{m^2}+\lambda_{\psi}\,.
\end{equation}
This might lead to numerical instabilities for large or diverging $\lambda_{\psi}$. The simple picture that the divergence of $\lambda_{\psi,\text{eff}}$ is connected to the onset of a nonvanishing expectation value for the bosonic field $\phi_0$, at least on intermediate scales, would not hold anymore. Furthermore, the dependence of the box-diagrams on the center of mass momentum would be neglected completely by this procedure. Close to the resonance the momentum dependence of the effective four-fermion interaction in the bosonized language as in Eq. (\ref{eq:lambdapsieff2}) is crucial, and this might also be the case for the particle-hole contribution.

Another, much more elegant way to incorporate the effect of particle-hole fluctuations is provided by the method of bosonization \cite{Gies:2001, Pawlowski}. For this purpose, we use scale dependent fields in the average action. The scale dependence of $\Gamma_k[\chi_k]$ is modified by a term reflecting the $k$-dependence of the argument $\chi_k$ \cite{Gies:2001}

\begin{equation}\label{eq:scalefield}
 \partial_k \Gamma_k[\chi_k]=\int\frac{\delta \Gamma_k}{\delta \chi_k}\partial_k\chi_k+\frac{1}{2}\mathrm{STr}\left[ \left(\Gamma_k^{(2)}+R_k \right)^{-1} \partial_k R_k\right] \,.
\end{equation}

For our purpose it is sufficient to work with scale dependent bosonic fields $\bar\phi$ and keep the fermionic field $\psi$ scale independent. In practice, we employ bosonic fields $\bar\phi_k^*$, and $\bar\phi_k$ with an explicit 
scale dependence which reads in momentum space
\begin{eqnarray}
\nonumber
\partial_k \bar \phi_k(q) & = & (\psi_1\psi_2)(q) \partial_k \upsilon\,,\\
\partial_k \bar \phi_k^*(q) & = & (\psi_2^\dagger\psi_1^\dagger)(q) \partial_k \upsilon.
\label{eq:scaledependenceoffields}
\end{eqnarray}
In consequence, the flow equations in the symmetric regime get modified

\begin{eqnarray}
 \partial_k \bar{h} &=& \partial_k \bar{h}{\big |}_{\bar \phi_k}-\bar{P}_{\phi}(q)\partial_k \upsilon\,,\\
 \partial_k \lambda_{\psi} &=& \partial_k \lambda_{\psi}{\big |}_{\bar \phi_k}-2\bar{h}\partial_k \upsilon.
\label{eq:modifiedflowequations}
\end{eqnarray}
Here $q$ is the center of mass momentum of the scattering fermions. In the notation of Eq. \eqref{eq:lambdapsieff} we have $q=p_1+p_2$ and we will take $\vec q=0$, and $q_0=0$. The first term on the right hand side in Eq. \eqref{eq:modifiedflowequations} gives the contribution of the flow equation which is valid for fixed field $\bar \phi_k$. The second term comes from the explicit scale dependence of $\bar \phi_k$. The inverse propagator of the complex boson field $\bar{\phi}$ is denoted by $\bar{P}_{\phi}(q)=\bar A_\phi P_\phi(q)=\bar A_\phi (m^2+i Z_\phi q_0+\vec q^2/2)$, cf. Eq. \eqref{eq:Bosonpropagator}. 

We can choose $\partial_k \upsilon$ such that the flow of the coupling $\lambda_{\psi}$ vanishes, i.e. that we have $\lambda_{\psi}=0$ on all scales. This modifies the flow equation for the renormalized Yukawa coupling according to

\begin{equation}
 \partial_k h = \partial_k h{\big |}_{\bar \phi_k}-\frac{m^2}{2h}\partial_k \lambda_{\psi}{\big |}_{\bar \phi_k}\,,
 \label{eq:modfiedflowofh}
\end{equation}
with $\partial_k h{\big |}_{\bar \phi_k}$ the contribution without bosonization and $\partial_k \lambda_\psi{\big |}_{\bar \phi_k}$ given by the box diagram in Fig. \ref{fig:boxes}. Since $\lambda_\psi$ remains zero during the flow, the effective four-fermion interaction $\lambda_{\psi,\text{eff}}$ is now purely given by the boson exchange. However, the contribution of the particle-hole exchange diagrams is incorporated via the second term in Eq. \eqref{eq:modfiedflowofh}. 

In the regime with spontaneously broken symmetry we use a real basis for the bosonic field
\begin{equation}
 \bar{\phi}=\bar{\phi}_0+\frac{1}{\sqrt{2}}(\bar{\phi}_1+i\bar{\phi}_2),
\end{equation}
where the expectation value $\bar{\phi}_0$ is chosen to be real without loss of generality. The real fields $\bar \phi_1$ and $\bar \phi_2$ then describe the radial and the Goldstone mode, respectively. To determine the flow equation of $\bar{h}$, we use the projection description

\begin{equation}\label{eq:projectiononh}
 \partial_k \bar{h}=i\sqrt{2}\Omega^{-1}\frac{\delta}{\delta\phi_2(0)}\frac{\delta}{\delta\psi_1(0)}\frac{\delta}{\delta\psi_2(0)}\partial_k \Gamma_k\,,
\end{equation}
with the four volume $\Omega=\frac{1}{T}\int_{\vec{x}}$. Since the Goldstone mode has vanishing ``mass'', the flow of the Yukawa coupling is not modified by the box diagram (Fig. \ref{fig:boxes}) in the regime with spontaneous symmetry breaking.

We emphasize that the non-perturbative nature of the flow equations for the various couplings provides for a resummation similar to the one in Eq. \eqref{PHComp}, and thus goes beyond the treatment by Gorkov and Melik-Barkhudarov \cite{Gorkov} which includes the particle-hole diagrams only in a perturbative way. Furthermore, the inner bosonic lines $h^2/P_\phi(q)$ in the box-diagrams represent the center of mass momentum dependence of the four-fermion vertex. This center of mass momentum dependence is neglected in Gorkov's pointlike treatment, and thus represents a further improvement of the classic calculation. Actually, this momentum dependence becomes substantial -- and should not be neglected in a consistent treatment -- away from the BCS regime where the physics of the bosonic bound state sets in. Finally, we note that the truncation \eqref{eq:truncation} supplemented with \eqref{eq:fourfermionvertex} closes the truncation to fourth order in the fields except for a fermion-boson vertex $\lambda_{\psi\phi}\psi^\dagger\psi\phi^*\phi$ which plays a role for the scattering physics deep in the BEC regime \cite{DKS} but is not expected to have an important impact on the critical temperature in the unitarity and BCS regimes.

\section{Critical temperature}\label{sec:CriticalTemperature}
To obtain the flow equations for the running couplings of our truncation Eq. \eqref{eq:truncation} we use projection prescriptions similar to Eq. \eqref{eq:projectiononh}. The resulting system of ordinary coupled differential equations is then solved numerically for different chemical potentials $\mu$ and temperatures $T$. For temperatures sufficiently small compared to the Fermi temperature $T_F=(3\pi^2n)^{2/3}$, $T/T_F\ll 1$ we find that the effective potential $U$ at the macroscopic scale $k=0$ develops a minimum at a nonzero field value $\rho_0>0$, $\partial_\rho U(\rho_0)=0$. The system is then in the superfluid phase. For larger temperatures we find that the minimum is at $\rho_0=0$ and that the ``mass parameter'' $m^2$ is positive, $m^2=\partial_\rho U(0)>0$. The critical temperature $T_c$ of this phase transition between the superfluid and the normal phase is then defined as the temperature where one has
\begin{equation}
\rho_0=0,\quad \partial_\rho U(0)=0\quad \text{at} \quad k=0.
\end{equation}
Throughout the whole crossover the transition $\rho_0\to0$ is continuous as a function of $T$ demonstrating that the phase transition is of second order.

\begin{figure}[ht]
 \centering
	\includegraphics[width=0.45\textwidth]{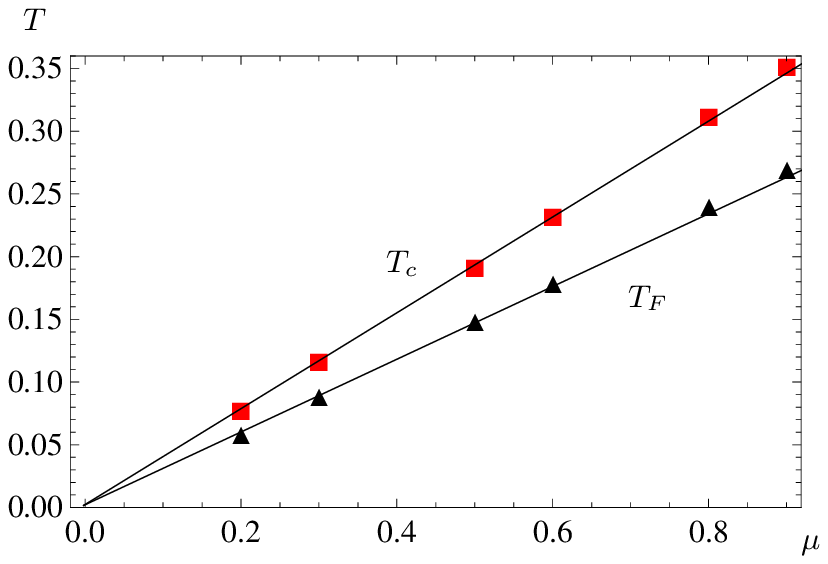}
	\caption{(Color online) Critical temperature $T_c$  (boxes) and Fermi temperature $T_F=(3\pi^2 n)^{2/3}$ (triangles) as a function of the chemical potential $\mu$. For convenience the Fermi temperature is scaled by a factor 1/5. We also plot the linear fits $T_c=0.39\mu$ and $T_F=1.48\mu$. The units are arbitrary and we use $\Lambda=e^7$.}
	\label{fig:tcrit2}
\end{figure}

In Fig. \ref{fig:tcrit2} we plot our result obtained for the critical temperature $T_c$ and the Fermi temperature $T_F$ as a function of the chemical potential $\mu$ at the unitarity point with $a^{-1}=0$. From dimensional analysis it is clear that both dependencies are linear, $T_c, T_F\sim \mu$, provided that non-universal effects involving the ultraviolet cutoff scale $\Lambda$ can be neglected. That this is indeed found numerically can be seen as a nontrivial test of our approximation scheme and the numerical procedures as well as the universality of the system. Dividing the slope of both lines gives $T_c/T_F=0.264$, a result that will be discussed in more detail below. 
We emphasize that part of the potential error in this estimates is due to uncertainties in the precise quantitative determination of the density or $T_F$.

\section{Phase diagram}\label{sec:PhaseDiagram}
\begin{figure}[ht]
\centering
\includegraphics[width=0.45\textwidth]{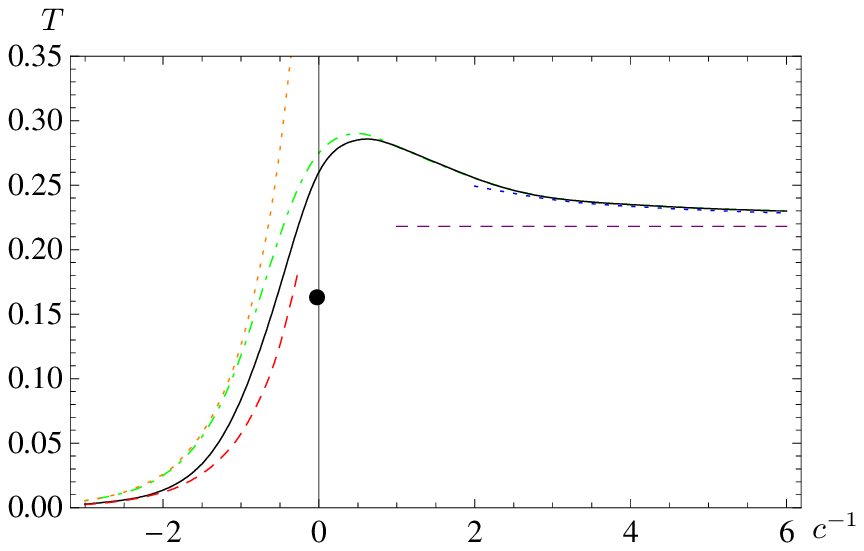}
\caption{(Color online) Dimensionless critical temperature $T_c/T_F$ as a function of the inverse concentration $c^{-1}=(a k_F)^{-1}$. The black solid line includes the effect of particle-hole fluctuations. We also show the result obtained when particle-hole fluctuations are neglected (dot-dashed line). For comparison, we plot the BCS result without (left dotted line) and with Gorkov's correction (left dashed). On the BEC side with $c^{-1}>1$ we show the critical temperature for a gas of free bosonic molecules (horizontal dashed line) and a fit to the shift in $T_c$ for interacting bosons, $\Delta T_c\sim c$ (dotted line on the right). The black solid dot gives the QMC results \cite{MonteCarloBuro,MonteCarloBulgac}.}
\label{fig:tcrit}
\end{figure}

The effect of the particle-hole fluctuations shows most prominently in the result for the critical temperature. With our approach we can compute the critical temperature for the phase transition to superfluidity throughout the crossover. The results are shown in Fig. \ref{fig:tcrit}. We plot the critical temperature in units of the Fermi temperature $T_c/T_F$ as a function of the scattering length measured in units of the inverse Fermi momentum, i.~e. the concentration $c=a k_F$.
We can roughly distinguish three different regimes. On the left side, where $c^{-1}\lesssim-1$, the interaction is weakly attractive. Mean field or BCS theory is qualitatively valid here. In Fig. \ref{fig:tcrit} we denote the BCS result by the dotted line on the left ($c^{-1}<0$). However, the BCS approximation has to be corrected by the effect of particle-hole fluctuations, which lower the value for the critical temperature by a factor of $2.2$. This is the Gorkov correction (dashed line on the left side in Fig. \ref{fig:tcrit}). The second regime is found on the far right side, where the interaction again is weak, but now we find a bound state of two atoms. In this regime the system exhibits Bose-Einstein condensation of molecules as the temperature is decreased. The dashed horizontal line on the right side shows the critical temperature of a free Bose-Einstein condensate of molecules. In-between there is the unitarity regime, where the two-atom scattering length diverges ($c^{-1} \rightarrow 0$) and we deal with a system of strongly interacting fermions.

Our result including the particle-hole fluctuations is given by the solid line. This may be compared with a functional renormalization flow investigation without including particle-hole fluctuations (dot-dashed line) \cite{Diehl:2007th}. For $c\to 0_-$ the solid line of our result matches the BCS theory including the correction by Gorkov and Melik-Barkhudarov \cite{Gorkov},
\begin{equation}
\frac{T_c}{T_F}=\frac{e^C}{\pi}\left(\frac{2}{e}\right)^{7/3} e^{\pi/(2c)}\approx 0.28 e^{\pi/(2c)}.
\end{equation}
In the regime $c^{-1}>-2$ we see that the non-perturbative result given by our RG analysis deviates from Gorkov's result, which is derived in a perturbative setting. 

On the BEC-side for very large and positive $c^{-1}$  our result approaches the critical temperature of a free Bose gas where the bosons have twice the mass of the fermions $M_B=2M$. In our units the critical temperature is then
\begin{equation}
\frac{T_{c,\text{BEC}}}{T_F}=\frac{2\pi}{\left[6\pi^2 \zeta(3/2)\right]^{2/3}}\approx 0.218.
\end{equation} 
For $c\to 0_+$ this value is approached in the form
\begin{equation}
\frac{T_c-T_{c,\text{BEC}}}{T_{c,\text{BEC}}}=\kappa a_M n_M^{1/3}=\kappa \frac{a_M}{a}\frac{c}{(6\pi^2)^{1/3}}.
\end{equation}
Here, $n_M=n/2$ is the density of molecules and $a_M$ is the scattering length between them. For the ratio $a_M/a$ we use our result $a_M/a=0.718$ obtained from solving the flow equations in vacuum, i.~e. at $T=n=0$, see also \cite{Diehl:2007th}. This has to be compared to the result obtained from solving the corresponding Schr\"odinger equation which gives $a_M/a=0.6$ \cite{Petrov}. For the coefficients determining the shift in $T_c$ compared to the free Bose gas we find $\kappa=1.55$. 

For $c^{-1}\gtrsim0.5$ the effect of the particle-hole fluctuations vanishes. This is expected since the chemical potential is now negative $\mu<0$ and there is no Fermi surface any more. Because of that there is no difference between the new curve with particle-hole fluctuations (solid in Fig. \ref{fig:tcrit}) and the one obtained when particle-hole contributions are neglected (dot-dashed in Fig. \ref{fig:tcrit}). Due to the use of an optimized cutoff scheme and a different computation of the density our results differ slightly from the ones obtained in \cite{Diehl:2007th}.

In the unitary regime ($c^{-1}\approx 0$) the particle-hole fluctuations still have a quantitative effect. We can give an improved estimate for the critical temperature at the resonance ($c^{-1}=0$) where we find $T_c/T_F=0.264$. Results from quantum Monte Carlo simulations are $T_c/T_F = 0.15$ \cite{MonteCarloBuro,MonteCarloBulgac} and $T_c/T_F = 0.245$ \cite{MonteCarloAkki}. The measurement by Luo \textit{et al.} \cite{Luo2007} in an optical trap gives $T_c/T_F = 0.29 (+0.03/-0.02)$, which is a result based on the study of the specific heat of the system.

\section{Crossover to narrow resonances}\label{sec:CrossoverNarrowResonances}
Since we use a two channel model (Eq. \eqref{eqMicroscopicAction}) we can not only describe broad resonances with $h_\Lambda^2\to \infty$ but also narrow ones with $h_\Lambda^2\to0$. This corresponds to a nontrivial limit of the theory which can be treated exactly \cite{Diehl:2005an, Gurarie2007}. In the limit $h_\Lambda\to 0$ the microscopic action Eq. \eqref{eqMicroscopicAction} describes free fermions and bosons. The essential feature is, that they are in thermodynamic equilibrium so that they have equal chemical potential. (There is a factor 2 for the bosons since they consist of two fermions.) For vanishing Yukawa coupling $h_\Lambda$ the theory is Gaussian and the macroscopic propagator equals the microscopic propagator. There is no normalization of the ``mass''-term $m^2$ so that the detuning parameter in Eq. \eqref{eqMicroscopicAction} is $\nu=\mu_M(B-B_0)$ and
\begin{equation}
m^2=\mu_M(B-B_0)-2\mu.
\end{equation}

To determine the critical density for fixed temperature, we have to adjust the chemical potential $\mu$ such that the bosons are just at the border to the superfluid phase. For free bosons this implies $m^2=0$ and thus
\begin{equation}
\mu=\frac{1}{2}\mu_M (B-B_0)=-\frac{1}{16\pi}h^2 a^{-1}.
\label{eq:ChemicalpotetialatTc}
\end{equation}
In the last equation we use the relation between the detuning and the scattering length
\begin{equation}\label{eq:detuningandscatlenth}
a=-\frac{h^2}{8\pi \mu_M(B-B_0)}.
\end{equation}
The critical temperature $T_c$ is now determined from the implicit equation
\begin{equation}
\int \frac{d^3 p}{(2\pi)^3}\left\{\frac{2}{e^{\frac{1}{T_c}(\vec p^2-\mu)}+1}+\frac{2}{e^{\frac{1}{2T_c}\vec p^2}-1}\right\}=n.
\label{eq:TcNarrowresonanceimplicit}
\end{equation}

\begin{figure}[ht]
\centering
\includegraphics[width=0.45\textwidth]{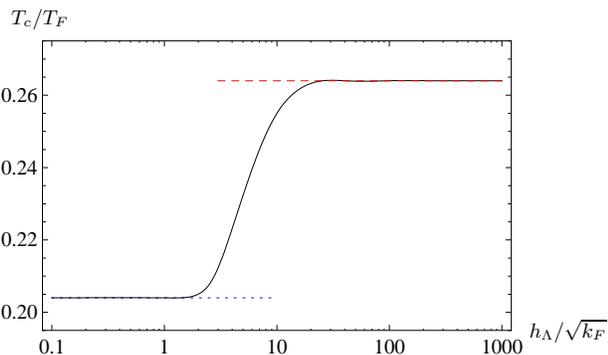}
\caption{(Color online) The critical temperature divided by the Fermi temperature $T_c/T_F$ as a function of the dimensionless Yukawa coupling $h_\Lambda/\sqrt{k_F}$ for $c^{-1}=0$ (solid line). One can clearly see the plateaus in the narrow resonance limit ($T_c/T_F \approx 0.204$, dotted line) and in the broad resonance limit ($T_c/T_F \approx 0.264$, dashed line).}
\label{fig:narrowbroad}
\end{figure}

While the BCS-BEC crossover can be studied as a function of $B-B_0$ or $\mu$, Eq. \eqref{eq:detuningandscatlenth} implies that for $h_\Lambda^2\to 0$ a finite scattering length $a$ requires $B\to B_0$. For all $c\neq 0$ the narrow resonance limit implies for the phase transition $B=B_0$ and therefore $\mu=0$. (A different concentration variable $c_\text{med}$ was used in \cite{Diehl:2005an, Diehl:2005ae}, such that the crossover could be studied as a function of $c_\text{med}$ in the narrow resonance limit, see the discussion at the end of this section.) For $\mu=0$ Eq. \eqref{eq:TcNarrowresonanceimplicit} can be solved analytically and gives
\begin{equation}
\frac{T_c}{T_F}=\left(\frac{4\sqrt{2}}{3(3+\sqrt{2})\pi^{1/2}\zeta(3/2)}\right)^{2/3}\approx 0.204.
\end{equation} 
This result is confirmed numerically by solving the flow equations for different microscopic Yukawa couplings $h_\Lambda$ and taking the limit $h_\Lambda\to 0$. In Fig. \ref{fig:narrowbroad}, we show the critical temperature $T_c/T_F$ as a function of the dimensionless Yukawa coupling $h_\Lambda/\sqrt{k_F}$ in the ``unitarity limit'' $c^{-1}=0$ (solid line). For small values of the Yukawa coupling, $h_\Lambda/\sqrt{k_F} \lesssim 2$ we enter the regime of the narrow resonance limit and the critical temperature is independent of the precise value of $h_\Lambda$. The numerical value matches the analytical result $T_c/T_F\approx 0.204$ (dotted line in Fig. \ref{fig:narrowbroad}). For large Yukawa couplings, $h_\Lambda/\sqrt{k_F} \gtrsim 40$, we recover the result of the broad resonance limit as expected. In between there is a smooth crossover of the critical temperature from narrow to broad resonances.

We use here a definition of the concentration $c=a k_F$ in terms of the vacuum scattering length $a$. This has the advantage of a straightforward comparison with experiment since $a^{-1}$ is directly related to the detuning of the magnetic field $B-B_0$, and the ``unitarity limit'' $c^{-1}=0$ precisely corresponds to the peak of the resonance $B=B_0$. However, for a nonvanishing density other definitions of the concentration parameter are possible, since the effective fermion interaction $\lambda_{\psi,\text{eff}}$ depends on the density. For example, one could define for $n\neq 0$ a ``in medium scattering length'' $\bar a=\lambda_{\psi,\text{eff}}/(8\pi)$, with $\lambda_{\psi,\text{eff}}=-h^2/m^2$ evaluated for $T=0$ but $n\neq 0$ \cite{Diehl:2005an}. The corresponding ``in medium concentration'' $c_\text{med}=\bar a k_F$ would differ from our definition by a term involving the chemical potential, resulting in a shift of the location of the unitarity limit if the latter is defined as $c_\text{med}^{-1}=0$. While for broad resonances both definitions effectively coincide, for narrow resonances a precise statement how the concentration is defined is mandatory when aiming for a precision comparison with experiment and numerical simulations for quantities as $T_c/T_F$ at the unitarity limit. For example, defining the unitarity limit by $c_\text{med}^{-1}=0$ would shift the critical temperature in the narrow resonance limit to $T_c/T_F=0.185$ \cite{Diehl:2005an}.

\section{Conclusion}\label{sec:Conclusions}
This paper aims at a quantitatively reliable computation of the phase diagram for the transition to superfluidity for non-relativistic fermions. We obtain an estimate of the critical temperature $T_c$ for the whole range of scattering length in the BCS-BEC crossover. A quantitatively precise determination of $T_c$ for the BCS-BEC crossover is a challenge both for theory and experiment. A successful match would extend the success of the ``universal critical physics'' to quantities that are ``nonuniversal'' in the language of critical phenomena as the critical temperature itself. On the theoretical side this requires a full control of the mapping from microphysics to macrophysics. It constitutes a decisive test for non-perturbative methods in complex many-body theories. New extended notions of universality become visible, which also extend to physics away from the critical temperature, as the low temperature limit $T\to 0$.

For non-relativistic fermions the task of a precise computation of $T_c$ is difficult even for small scattering length. The basic reason is that the phase transition itself is a non-perturbative phenomenon, linked to an effective fermion interaction $\lambda_{\psi,\text{eff}}$ growing to large values. In order to locate $T_c$ precisely, one has to follow the scale-dependence of $\lambda_{\psi,\text{eff}}$ with sufficient precision. The problem arises from the substantial momentum dependence of the fluctuation contributions to the effective four-fermion vertex. In principle, a consistent one loop approximation to the flow should follow the full momentum dependence of the vertex according to Eq. \eqref{eq:lambdapsi2}. With the general form $T_c/T_F=F e^{\pi/(2k_F a)}$ for $a k_F\to 0_-$ this will influence the value of the prefactor $F$. Different projections of the flow  in Eq. \eqref{eq:lambdapsi2} onto a single coupling $\lambda_\psi$, as the BCS approximation or the Gorkov projection prescription, result in different approximations for $F$.

In this paper we use a projection that coincides with the Gorkov approximation in the BCS-limit $a k_F\to 0_-$. Our method of non-perturbative flow equations allows us to extend Gorkov's result to the whole range of $a$, including the ``unitarity limit'' $a\to \infty$. Already at the present stage of truncation our result $T_c/T_F=0.264$ for $a\to\infty$ agrees well with experiment and competes with quantum Monte-Carlo simulations.  Further extensions of the truncation should refine the estimate of $T_c$ and also permit an indication of the theoretical uncertainty of this estimate.

\section*{Acknowledgement}
We thank H. Gies, H.~C.~Krahl, J.~M.~Pawlowski, R.~Schmidt and P.~Strack for useful discussion. M.~S. acknowledges support by the DFG under Contract No. Gi~328/1-4 (Emmy-Noether Program) and by the DFG research unit FOR723 under Contract No. We~1056/9-1.

\end{document}